\documentclass[twocolumn, a4paper]{article}
\usepackage{0_arxiv}

\usepackage[utf8]{inputenc} 
\usepackage[T1]{fontenc}    
\usepackage{hyperref}       
\usepackage{url}            
\usepackage{booktabs}       
\usepackage{amsfonts}       
\usepackage{nicefrac}       
\usepackage{microtype}      
\usepackage{fancyhdr}       
\usepackage{lipsum}
\usepackage[ddmmyyyy]{datetime}
\newdateformat{mydate}{\twodigit{\THEDAY}{ }\shortmontfsubfihname[\THEMONTH], \THEYEAR}

\usepackage{amsmath} 
\usepackage{amssymb} 
\usepackage{dsfont}
\usepackage[pdftex,dvipsnames]{xcolor}
\usepackage[colorinlistoftodos,prependcaption]{todonotes}

\usepackage[babel,german=guillemets]{csquotes}
\usepackage{float}
\usepackage{graphicx} 
\usepackage{listings}
\usepackage{algorithm}
\usepackage[noend]{algpseudocode}
\usepackage{makecell}
\usepackage{microtype} 
\usepackage{mparhack}
\usepackage{multicol} 
\usepackage{multirow} 
\usepackage{mathtools}
\usepackage{paralist} 
\usepackage{wrapfig} 
\usepackage{physics} 
\usepackage[verbose]{placeins}
\usepackage{siunitx}
\usepackage[format=plain, font={small,up}, labelfont=bf, justification=justified, labelsep=period, skip=10pt]{caption}

\usepackage{subcaption}
\usepackage[nottoc]{tocbibind}
\usepackage{my_newcommands}
\usepackage{algpseudocode}
\usepackage{algorithm}
\usepackage{cooltooltips}
\usepackage{wrapfig}
\usepackage{orcidlink}

\title{Ultra-directional and high-efficiency µLEDs via gradient index filled micro-Horn collimators}

\author{
    Alexander Luce \orcidlink{0000-0002-9659-6579} \\
    University Erlangen-Nürnberg \&\\
    Max Planck Institute \\
    for the Science of Light \&\\
    ams-OSRAM \\
    Regensburg \\
\And
    Rasoul Alaee \orcidlink{0000-0003-2187-4949} \\
    ams-OSRAM\\
    Regensburg \\
\And
    Aimi Abass \\
    ams-OSRAM\\
    Regensburg \\
}

\begin{document}

\twocolumn[ 
  \begin{@twocolumnfalse} 
  
\maketitle

\begin{abstract}

Micro-LEDs (\textmu LEDs) are poised to transform AR/VR, display, and optical communication technologies, but they are currently hindered by low light extraction efficiency and non-directional emission. Our study introduces an innovative approach using a descending index multilayer anti-reflection coating combined with a horn collimator structure atop the \textmu LED pixel. This design leverages the propagation of light outside the critical angle to enhance both the directionality and extraction efficiency of emitted light. By implementing either discrete or continuous refractive index gradients within the horn, we achieve a dramatic tenfold increase in light extraction within a $\pm$15° cone, with an overall light extraction efficiency reaching approximately 80\%, where 31\% of the power is concentrated within this narrow cone. This performance surpasses that of an optimized SiO2 half-ellipsoidal lens, which diameter and height is 24X and 26X larger than the pixel width respectively, while our design only slightly increases the device height and expands the final light escape surface to 3 times and roughly 4 times the pixel width respectively. Such efficiency, directionality enhancement, and compactness make this solution particularly suitable for high-resolution, densely packed \textmu LED arrays, promising advancements in high-performance, miniaturized display systems.
\end{abstract}
\vspace{0.35cm}

  \end{@twocolumnfalse} 
]


\section{Introduction}
\label{sec:1_introduction}

Generating light with high efficiency and directionality directly from the light source itself is in general a highly desirable characteristic for LEDs and for µLEDs \cite{Taki:2019:Visible_LEDs:More_than_efficient_light}, in particular for AR/VR and display applications \cite{Xiong:2021:Augmented_reality_and_virtual_reality_displays:_emerging_technologies_and_future_perspectives, Kang:2024:Advances_in_display_technology:_augmented_reality_virtual_reality_quantum_dot-based_light-emitting_diodes_and_organic_light-emitting_diodes} or optical communications \cite{Lu:2022:High-speed_visible_light_communication_based_on_micro-LED:_A_technology_with_wide_applications_in_next_generation_communication, Singh:2020:Micro-LED_as_a_Promising_Candidate_for_High-Speed_Visible_Light_Communication}. However, besides the difficulties in manufacturing and package integration of µLEDs \cite{Templier:2023:MicroLED_Technology:A_Unique_Opportunity_Toward_More_Than_Displays, Anwar:2022:Recent_Progress_in_Micro-LED-Based_Display_Technologies, Chen:2023:Integration_Technology_of_Micro-LED_for_Next-Generation_Display} current µLEDs still offer significant room for improvement in terms of efficiency as suggested by the external quantum efficiency (EQE) of µLEDs which is typically still only in the range of few percent \cite{Baek:2023:Ultra-low-current_driven_InGaN_blue_micro_light-emitting_diodes_for_electrically_efficient_and_self-heating_relaxed_microdisplay, Li:2023:Significant_Quantum_Efficiency_Enhancement_of_InGaN_Red_Micro-Light-Emitting_Diodes_with_a_Peak_External_Quantum_Efficiency_of_up_to_6}. 

The EQE measures how much from the electrical energy injected into the LED is converted to useful light outside of the LED. It consists of three contributions, the electrical internal quantum efficiency (IQE$_e$), the purcell factor (pf) and the total light extraction efficiency ($\eta_\text{LEE}$ = LEE$_{90}$). Often, purcell factor and electrical internal quantum efficiency is combined as $\eta_\text{IQE}$ \cite{Schubert:2006:Light-Emitting_Diodes}. For many applications, only the light into a particular solid angle $\Gamma$ in the farfield is of interest and therefore LEE$_\Gamma$ is given which also considers the directionality. The fraction of light which is emitted into the solid angle $\Gamma$ over the total solid angle is denoted by $\eta_\Gamma$. As a consequence, all of those factors determine the overall performance of the µLED, the so called wall-plug-efficiency ($\eta_\text{WPE}$), and need to be maximized to achieve the overall best device performance. In general, the total wall-plug-efficiency of a µLED is therefore a product of different contributions 
\begin{align*}
    \eta_\text{WPE} = \eta_\text{EL}\,\eta_\text{IQE}\,\eta_\text{LEE}\,\eta_\Gamma,
\end{align*}

where $\eta_\text{EL}$ denotes the efficiency of the electrical system. The IQE of µLEDs remains a big challenge due to defect recombination on the interface of the active region with its surroundings \cite{Vogl:2024:Optical_characteristics_of_thin_film-based_InGaN_micro-LED_arrays:_a_study_on_size_effect_and_far_field_behavior, Zhang:2023:Recent_Advances_on_GaN-Based_Micro-LEDs}. In particular for the smallest µLEDs, the surface-to-bulk ratio becomes increasingly skewed and therefore a strain on the IQE due to extrinsic and intrinsic surface defects \cite{Wang:2023:3.5×3.5μm_GaN_blue_micro-light-emitting_diodes_with_negligible_sidewall_surface_nonradiative_recombination}.

To enhance the general light extraction efficiency of a LED or µLED, typical approaches involve utilizing resonant/lossy cavities with distributed-bragg-reflector- (DBR) / anti-reflection-coatings \cite{Lee:2024:InGaN_blue_resonant_cavity_micro-LED_with_RGY_quantum_dot_layer_for_broad_gamut_efficient_displays, Wang:2023:GaN-on-Si_micro_resonant-cavity_light-emitting_diodes_with_dielectric_and_metal_mirrors} or introducing surface textures \cite{Li:2023:Significant_Quantum_Efficiency_Enhancement_of_InGaN_Red_Micro-Light-Emitting_Diodes_with_a_Peak_External_Quantum_Efficiency_of_up_to_6}. In particular anti-reflection-coatings can improve the direct outcoupling efficiency by reducing reflection losses at the interface. However, it is not possible to extract light with incidence angle beyond the total-internal-reflection (TIR) angle. DBRs can suppress emission at shallow angles but typically show higher losses due to absorption since the light interacts more often with absorbing materials inside the µLED pixel \cite{Schubert:2006:Light-Emitting_Diodes, Schubert:1992:Resonant_cavity_light-emitting_diode}. In applications where light within a particular narrow solid angle is useful, conventional rough outcoupling textures typically lead to a lambertian like far-field emission and therefore resulting in less useful light. However, if only the light into a particular solid angle is useful for the application, texturization typically leads to a lambertian-like farfield distribution and therefore low directionality which decreases the overall efficiency significantly. 

To reduce emission towards undesirable directions in the farfield, lot of effort is put into the directionality of the emission. Typical approaches involve utilizing micro-lens arrays \cite{Li:2023:Highly_efficient_and_ultra-compact_micro-LED_pico-projector_based_on_a_microlens_array}, surface texturing approaches \cite{Liu:2023:A_Review_on_Micro-LED_Display_Integrating_Metasurface_Structures}, meta-lenses \cite{Chen:2024:Broadband_beam_collimation_metasurface_for_full-color_micro-LED_displays}, µLED nanorods \cite{Veeramuthu:2024:Scalable_InGaN_nanowire_µ-LEDs:_paving_the_way_for_next-generation_display_technology, Wu:2022:InGaN_micro-light-emitting_diodes_monolithically_grown_on_Si:_achieving_ultra-stable_operation_through_polarization_and_strain_engineering} or photonic crystals \cite{Abdelkhalik:2023:Surface_lattice_resonances_for_beaming_and_outcoupling_green_µLEDs_emission, Chen:2023:Metamaterials_for_light_extraction_and_shaping_of_micro-scale_light-emitting_diodes:_from_the_perspective_of_one-dimensional_and_two-dimensional_photonic_crystals}. In particular for the smallest µLEDs which are also the most interesting for AR/VR applications, increasing directionality and forward efficiency is especially difficult due to size of the devices. Traditional optical elements such as lenses behave differently than for large LEDs due to near-field wave-optical properties of the µLED emission which necessitates larger devices overall. Nanorods and photonic crystal based architectures require multiple periods of the to effectively utilize the properties of the photonic bandstructure for directionality. Especially for the smallest devices, where the desired active region of a single pixel is in the range of the wavelength of the emitted light, this becomes infeasible since simply not sufficient periods can be realized in such a small size. For the smallest monolithic pixels, die shaping, the process of altering the geometry of the pixel itself to shape the light appropriately, has shown promising results for enhancing directionality and outcoupling efficiency \cite{Vogl:2023:Role_of_pixel_design_and_emission_wavelength_on_the_light_extraction_of_nitride-based_micro-LEDs, Vogl:2024:Optical_characteristics_of_thin_film-based_InGaN_micro-LED_arrays:_a_study_on_size_effect_and_far_field_behavior, Huang:2024:InGaN-based_blue_resonant_cavity_micro-LEDs_with_staggered_multiple_quantum_wells_enabling_full-color_and_low-crosstalk_micro-LED_displays, Park:2022:Interplay_of_sidewall_damage_and_light_extraction_efficiency_of_micro-LEDs, Chu:2023:Fabricating_and_investigating_a_beveled_mesa_with_a_specific_inclination_angle_to_improve_electrical_and_optical_performances_for_GaN-based_micro-light-emitting_diodes, Wang:2024:On_the_origin_of_the_enhanced_light_extraction_efficiency_of_DUV_LED_by_using_inclined_sidewalls}.

Meanwhile, the problem of directionality has been thoroughly investigated in the field of microwave radiation and antenna technology where antenna geometries exists with large directionality and high efficiency \cite{Balanis:2005:Antenna_Theory:_Analysis_and_Design}. However, there are significant differences between typical microwave antennas and µLEDs: 
\noindent
\begin{itemize}
    \item The wavelength of the radiation in µLEDs is small compared to the size of the device while the geometry is typically similar in size of the wavelength for microwaves applications.
    \item The emission for µLEDs originate from an active region which is large compared to the wavelength. 
    \item The emission coming from different parts of the active region is incoherent due to the spontaneous emission process of radiative recombination \cite{Schubert:2006:Light-Emitting_Diodes}. 
    \item µLEDs exhibit a much greater number of available and populated modes due to the distributed nature of the emission from the active region and the overall size of the device compared to the wavelength. 
    \item In microwave applications, metals can be considered to be perfect electric conductors while they have considerable penetration depth and absorption at optical wavelengths for µLEDs.
\end{itemize}

In this contribution, we investigate the application of a so called Horn collimator/antenna \cite{Balanis:2005:Antenna_Theory:_Analysis_and_Design} in enhancing light extraction efficiency and directionality of the emitted light. Although antenna theory assumes coherent emitters, we demonstrate here that the directionality and outcoupling enhancement is also seen for the case of incoherent emitters as in the case of \textmu LEDs. Furthermore, we demonstrate that the outcoupling and collimating performance of the horn antenna is improved by either introducing a discrete or continuously varying descending refractive index layer inside of the Horn. In the presence of such a layer, light from the \textmu LED, which would typically have momentum beyond the escape cone from semiconductor to air, can enter the horn collimator antenna structure. Within this GRIN layer, the pathway of such light is bent, causing a major portion of it to interact with the sidewalls, and is subsequently redirected into a more advantageous direction. In essence, adding the material complexity within the horn leads to the possibility of achieving strong directionality and large outcoupling efficiency enhancement with a smaller geometrical footprint, which is extremely desirable for high-resolution display applications. In describing the governing mechanisms that lead to outcoupling and directionality enhancement, we first show proof-of-principle 2D calculations. We then proceed to illustrate these principles through 3D cylindrical models, demonstrating the core findings from our 2D calculations. These models are compared against two pertinent reference scenarios: 1) a bare \textmu LED pixel, and 2) a \textmu LED pixel paired with a large, optimized SiO2 half-ellipse lens, where the lens's diameter is 40 times that of the \textmu LED pixel's active region and 24 times the pixel's opening width.

The results show an improvement in LEE$_\text{15}$ of up to one order of magnitude compared to the bare \textmu LED pixel and more than double compared to the \textmu LED pixel with the large SiO2 half-ellipsoidal lens. Achieving such an improvement in total LEE and LEE$_\Gamma$ with minimal geometrical size penalties for real devices would drastically change the prospects of applications of µLEDs in AR/VR and optical communication. 


\section{Theory}
\label{sec:theory}

\begin{figure*}[h]
  \centering
  \includegraphics[angle=0, trim = 0cm 0cm 0cm 0cm, clip, width = 1.0\textwidth]{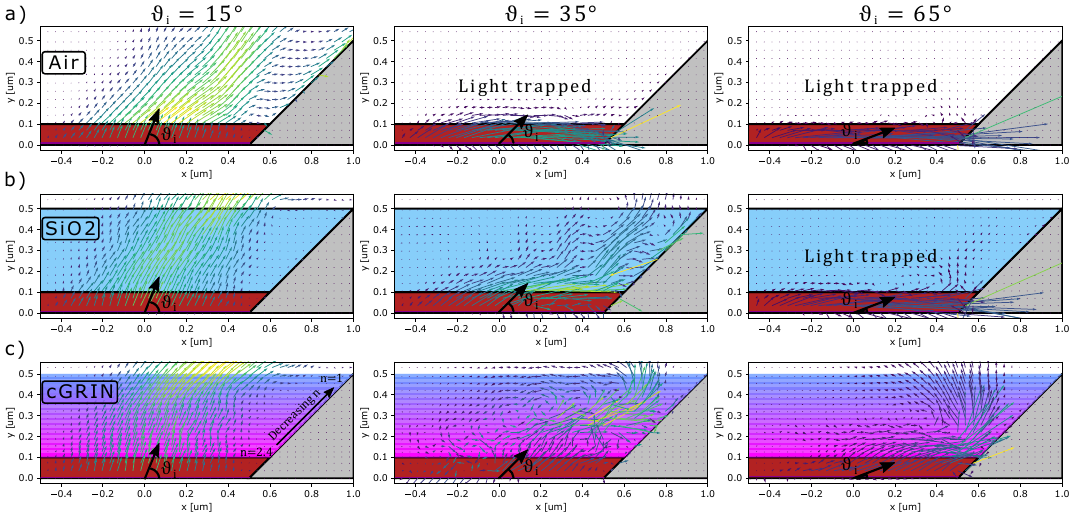}
  \caption{ \textbf{| Interaction of Gaussian Beams at different incident angles with the \textmu Horn and different filling materials.}
  2D model of a portion of the \textmu LED structure under investigation. The arrows indicate the power flow of a Gaussian beam launched from the semiconductor side at different angles as it propagates through the \textmu Horn collimator with different filling materials. Note that the model is symmetric around the y-axis but only a part of the geometry is displayed. A Gaussian beam with $\lambda=$ 450nm and injection angles [15°, 35°, 65°] is injected in different \textmu Horn filling media. Even without filling (row (a)), appropriate sidewalls increase the directionality of the light by redirecting light at the steepest emission angles with respect to the forward direction. Adding an additional filling of SiO2 (row (b)) improves outcoupling but still shows total internal reflection for more oblique angles. By adding a continuous step-down GRIN-style  (row (c)) inside the \textmu Horn antenna, much less of the light experiences total internal reflection and even more of the emitted light interacts with the sidewalls. Hence the LEE is increased much more than what would be possible via an anti-reflection coating. For the steepest injection angles, it becomes visible that even light which experiences total internal reflection is able to escape the device. In particular, the height h and sidewall angle $\vartheta$ of the \textmu Horn collimator influence how much of the injected power escapes the device and at which angle. \label{fig:concept_model_with_power_flow}}
\end{figure*}



As outlined in the Introduction, this study aims to investigate and enhance the light extraction efficiency (LEE) and directionality of a µLED, specifically focusing on the combined factor of $\eta_\text{LEE} \cdot \eta_\Gamma$. 

Fundamentally, an aperture antenna \cite{Balanis:2005:Antenna_Theory:_Analysis_and_Design} can be modeled as a Huygens source, which helps in understanding the farfield in terms of diffraction. It is well known that the diffraction of plane waves at a slit aperture results in an intensity distribution given by $I(\theta; \phi_i) = \text{sinc}\left(\frac{d \pi}{\lambda}(\sin{\theta} \pm \sin{\phi_i})\right)$ in the farfield, where $\phi_i$ is the incident angle on the aperture and $\lambda$ the wavelength of the light. Increasing the aperture diameter $d$ leads to a more directional farfield. Conversely, a broad incident near-field angular distribution causes a widening of the farfield due to the superposition effect caused by the incident angle. For better illustration, we demonstrate in \autoref{appendix:fig:huygens_source_diffraction} a wide near field angular distribution affects the directionality in the farfield.

Therefore, the emission characteristics of a µLED are inherently constrained by the diffraction of light at the outcoupling region, which acts as an aperture of limited size. More importantly, the light reaching the aperture typically has a broad angular distribution. The broad angular distribution is a consequence of the emission from the quantum wells, which couple to many modes within the µLED. This effect is particularly pronounced in solid-state µLEDs, where the semiconductor layer thickness and lateral size are several times the effective wavelength. Furthermore, these modes are broadened due to the lossy nature of the µLED cavity, which experiences both absorption and radiative losses, the latter, of course, being desirable. 

\begin{figure}[h]
  \centering
  \includegraphics[angle=0, trim = 0cm 0cm 0cm 0cm, clip, width = .99\columnwidth]{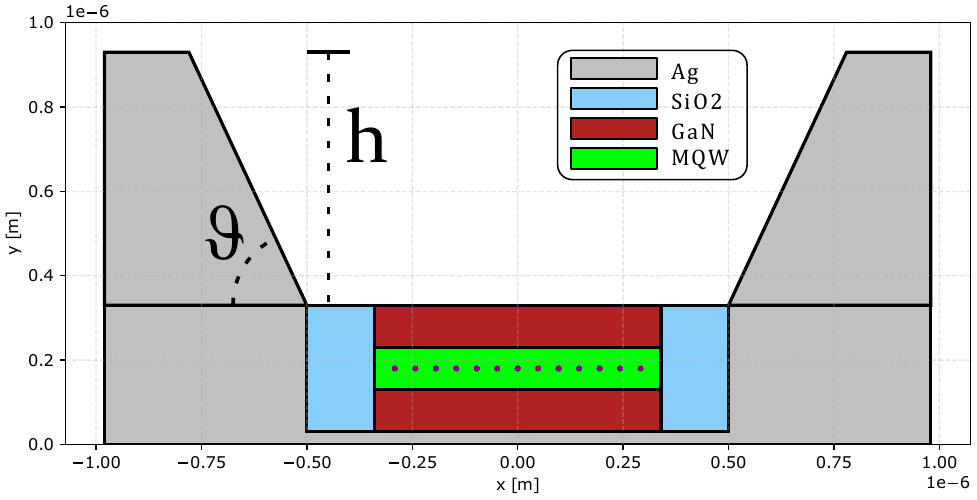}
  \caption{
  \textbf{| Schematic cross-section of the µLED model investigated in this work.} The \textmu Horn height h and the sidewall angle $\vartheta$ are systematically adjusted to determine the combination with the best LEE$_{15}$ for different \textmu Horn filling materials. The violet dots represent the position of the quantum well which is approximated by using a fixed amount of dipoles. From there, the incoherent superposition of the weighted emission from the dipoles is used to determine the LEE$_{15}$. More details are given in \autoref{sec:Effect of the Horn antenna on µLED emission}.
  \label{fig:sweep_parameters}}
\end{figure}

Although aperture size is typically application dependent, the angular distribution width can be controlled to some extent through the design of the µLED geometry. For example, managing interference effects with the rear mirror and/or incorporating slanted sidewalls in the µLED can influence this distribution. However, optimizing the sidewalls of a µLED for light extraction and directionality is often challenging. This is because slanted sidewalls, which cannot be properly passivated, may lead to a reduction in internal quantum efficiency. In this work, we propose to further increase the directionality of a µLED by utilizing a \textmu Horn collimator structure on top of the µLED, as shown in \autoref{fig:sweep_parameters} which avoids the necessity to modify the \textmu LED sidewalls.

It is crucial to consider that both the directionality of light and $\eta_\text{LEE}$ play a significant role. However, reducing the angular distribution with a steeper \textmu Horn could potentially impair LEE, necessitating a trade-off or a taller but less steep structure. In the following sections, we will explore how the different geometrical parameters which describe our \textmu Horn collimator: height, opening angle and material fillings, impacts the figure of merit of interest $\eta_\text{LEE}$ and $\eta_\Gamma$.

\subsection{GRIN and \textmu Horn interaction with a Gaussian beam}
In the simple ray optics picture, a \textmu Horn collimator works by redirecting oblique light propagating within the structure into a small angular cone around the normal direction through interaction with reflective slanted sidewalls. Assuming angularly isotropic emission entering the \textmu Horn collimator, a large geometry is typically needed to ensure that a significant portion of the power traveling in undesirable directions interacts with the slanted sidewalls and is redirected into the desired angular cone. Naturally, not all stray light will be redirected in the desired direction, especially if the target angular cone is small. To some extent, if the \textmu Horn collimator has a smoothly varying surface, the quality of collimation can be improved. However, in general, a large cone is required to significantly increase power in the desired direction.


When the \textmu Horn collimator is unfilled, containing only air, and especially when the \textmu LED pixel is not significantly smaller than the effective wavelength, a considerable amount of light that falls outside the escape cone from the semiconductor to air remains trapped within the \textmu LED (see \autoref{fig:concept_model_with_power_flow} (a)) where we show through 2D FDTD calculations the power flow of a Gaussian beam at 450 nm wavelength launched from the semiconductor side (GaN) to the interface of the micro-\textmu Horn structure at angles of 15°, 35°, and 65° from the normal). Since less light enters the \textmu Horn collimator, a larger opening angle and height are generally required to concentrate the limited light that can enter into the desired direction.

When the \textmu Horn is filled with a homogeneous dielectric material, more light from the \textmu LED can enter the \textmu Horn collimator because the escape cone from the semiconductor into the \textmu Horn expands (see \autoref{fig:concept_model_with_power_flow} (b)).
However, light within the semiconductor-to-air escape cone travels through the dielectric at less oblique angles, reducing interactions with the \textmu Horn's sidewalls and thus decreasing directionality. Conversely, light with momentum near the edge of the semiconductor-to-dielectric escape cone bends more sharply, interacting more with the \textmu Horn's sidewalls. This interaction redirects the light favorably, compensating for some loss of directionality. Typically, the MQW in the \textmu LED emits more light at heavily oblique angles outside the semiconductor-to-air escape cone. By filling the \textmu Horn collimator with a dielectric material, light from these angles can be redirected into the desired cone, although the potential for enhancement is limited because redirection is less effective for other momentum regions.

By introducing a gradient-index layer, either discrete (dGRIN) or continuous (cGRIN), we not only allow more light to enter the \textmu Horn collimator but also enable light with varying momenta to follow curved paths at different depths within the \textmu Horn. This interaction with the sidewalls results in more light being redirected into the desired cone. This effect is particularly evident in \autoref{fig:concept_model_with_power_flow} c), especially in the middle and right panels, where light outside the primary semiconductor-to-air escape cone, entering at various angles, is significantly bent at different depths and then redirected by the sidewall. Essentially, the power flow is bent within the gradient index until it starts propagating parallel to the interfaces. Just before the radiation starts being redirected downwards, it is intercepted by the side mirror and redirected upwards, rendering it useful for the application and thus increasing the directionality.


In theory, a continuous gradient index material ranging from a the maximum refractive index of the µLED down to the refractive index of the ambient medium should achieve the best results for light extraction and redirection. However, in the next sections we demonstrate that even with rough, discrete steps in refractive index, using naturally available materials, a significant enhancement in light outcoupling and directionality can be achieved. 

\begin{figure*}[h!]
  \centering
  \includegraphics[angle=0, trim = 0cm 0cm 0cm 0cm, clip, width = .99\textwidth]{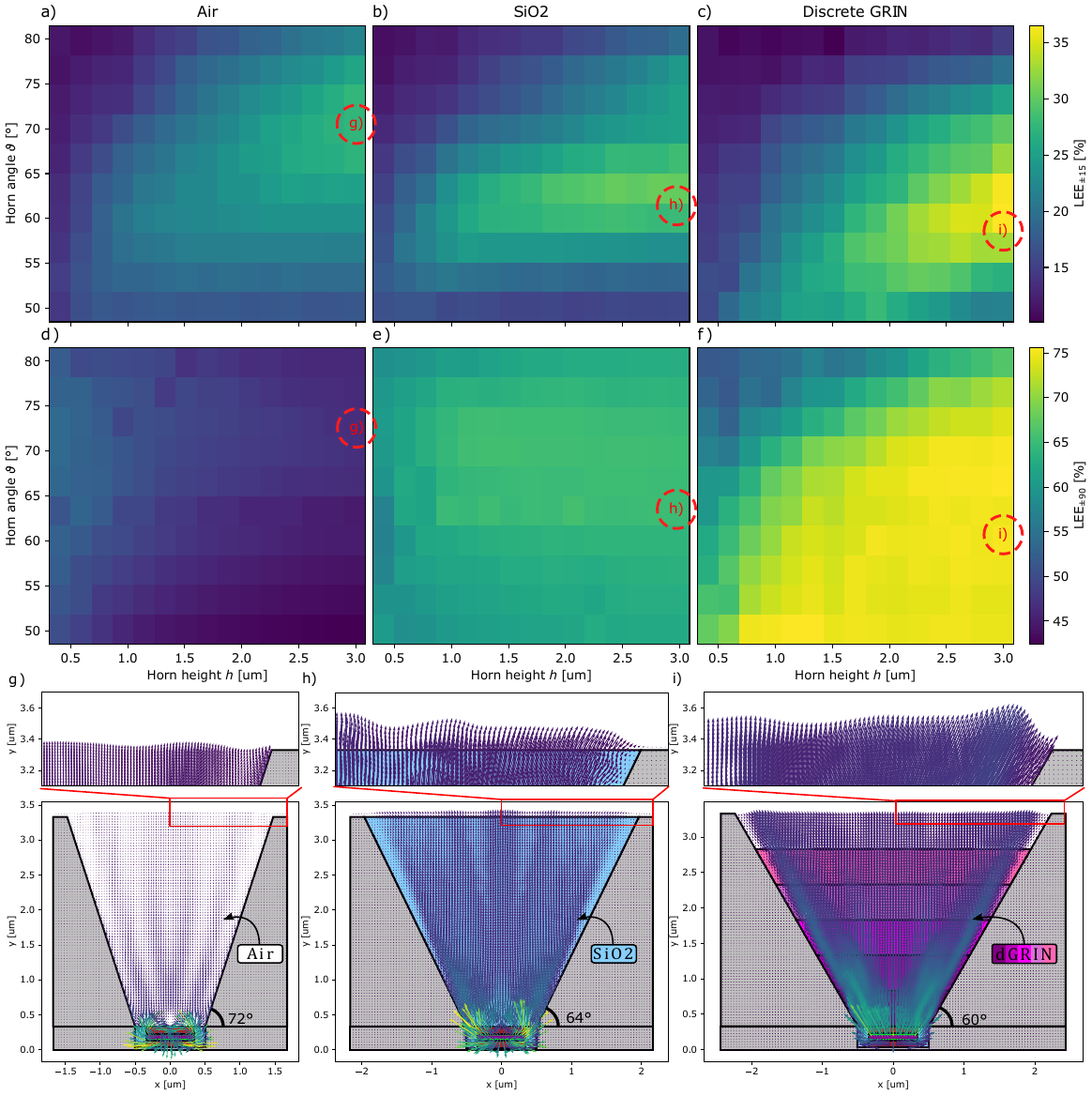}
  \caption{\textbf{| Power flow of the best \textmu Horn designs for different filler materials and the sweep results of \textmu Horn angle $\mathbf{\vartheta}$ and \textmu Horn height h of LEE$_{15}$.} In a), b), and c), the LEE$_{15}$ results are shown over the varying \textmu Horn height and opening angle for different \textmu Horn filling materials, while d), e), and f) show the results for LEE$_{90}$. a) and d) display the changing LEEs for a filling with air, b) and e) for SiO2, and c) and f) for the discrete GRIN. The best combinations of \textmu Horn height h and \textmu Horn angle $\vartheta$ are indicated, and the respective power flows are shown in g), h), and i). Evidently, the optimal \textmu Horn height shifts with respect to the height of the device, while the absolute directionality also increases with greater height of the \textmu Horn in all three cases. However, the optimal angle depends on the filler material. For the air-filled \textmu Horn, the best result is achieved with $\vartheta=72^\circ$ shown in g), for SiO2 at $\vartheta=64^\circ$ shown in h), and for the discrete GRIN with $\vartheta=60^\circ$ shown in i). Comparing the LEE$_{15}$ with the LEE$_{90}$ results in d), e), and f), it becomes evident that the filler material has a large impact on the total outcoupling efficiency, which contributes to the forward emission as well. \label{fig:2D_sweep_results}}
\end{figure*}

Obviously, the aforementioned effects only occur when the gradient refractive index layer is thick enough such that light does not simply perceive the entire layer as having a single effective index. The exact thickness at which all the desired effects occur can only be determined through rigorous calculations. Hence, in the following sections, we investigate the effect of the refractive index of various GRIN coatings and other filling materials with respect to the \textmu Horn size and angle.

\subsection{Effect of the \textmu Horn collimator on µLED emission}\label{sec:Effect of the Horn antenna on µLED emission}
To elucidate how the GRIN-filled horn collimator alters \textmu LED emission characteristics compared to traditional setups, we demonstrate via rigorous 2D FDTD simulations how the light within the $\pm$15° cone is affected by variations in horn height and angle across three different dielectric filling scenarios. The first one is air, the second SiO2, and the third one the GRIN filling which has a discretely varying index from that of GaN to air. Hence, we investigate a simplified µLED structure as shown in \autoref{fig:sweep_parameters} with different materials inside the \textmu Horn. The emission from the quantum well is approximated with dipoles emitting at different positions along the central plane of the active region. For simplicity, the dipole polarization is assumed to be isotropic.

In the next step. see \autoref{sec:3d_results}, we present 3D calculations for cylindrical structures at selected parameters to demonstrate that the enhancements observed in these proof-of-principle calculations are also visible in 3D and even more so due to the fact that more light are emitted in oblique directions beyond the escape cone by point dipole sources as compared to line dipole sources.

The toy µLED model consists of a simplified semiconductor epitaxial stack for the \textmu LED, comprised of a homogeneous layer of GaN. As seen in \autoref{fig:sweep_parameters}, the semiconductor thickness is 300 nm and has a width of 600 nm. We further consider that the \textmu LED is placed on top of a silver substrate which envelops the p- and n-sides of the µLED. On the sides, a SiO2 layer is placed, serving as electrical passivation with a width of 200 nm. The semiconductor region and the sidewall passivation layer thus effectively make a total width of 1 µm, which we consider as the main pixel width where light can escape from (ignoring the metal sidewall thicknesses). Finally, on top of the structure, the \textmu Horn collimator is placed. For simplicity, the horn collimator reflector is also considered to be comprised of silver.

To gain the most insights into the behavior of the \textmu Horn, we investigate three different but realistic scenarios: 
\begin{itemize}
    \item In the first scenario, the \textmu Horn is filled with air. 
    \item In the second scenario, it is filled with GaN to match the refractive index of the quantum wells.
    \item In the third scenario, we implement a discrete GRIN, which is equivalent to a multilayer stack of descending index by using a six-layer step-down coating composed of Nb2O5, Si3N4, Al2O3, SiO2, MgF2, and air, starting with Nb2O5.
\end{itemize}

For the study, we sweep the \textmu Horn angle from $\vartheta$ [deg] $= [50,\,80]$  and the height h [µm] $= [0.4,\, 3.0]$. 
To compute the LEE$_\Gamma$ we first compute the total emitted farfield by the \textmu LED pixel $\rho_\text{LEE}$ in 2D
\begin{align}
    \rho_\text{LEE}(\theta, \lambda) = \frac{n\, c\, \varepsilon_0}{2 P_0(\lambda) M}  \sum_{i=1}^M w_{i, \lambda, \text{pol}} \left| \mathbf{E}_i(\theta, \lambda)  \right|^2
\end{align}
and 3D
\begin{align}
    \rho_\text{LEE}(\theta, \varphi, \lambda) = \frac{n\, c\, \varepsilon_0}{2 P_0(\lambda) M}  \sum_{i=1}^M w_{i, \lambda, \text{pol}} \left| \mathbf{E}_i(\theta, \varphi, \lambda)  \right|^2
\end{align}
where $P_0(\lambda)$ is the energy injected into the simulation, $M$ is the number of considered dipoles and $w_{i, \lambda, \text{pol}}$ a weighting factor, $n$ the refractive index of the ambient medium, $c$ the speed of light and $\varepsilon_0$ the vacuum permittivity. This weighting factor accounts for a dipole position dependent weighting, the spectrum of the emission, and a correction for the contribution to different polarization. For 2D simulations, the position dependence of dipole emitters does not play a role because the translational symmetry introduces a uniform weighting over the active area. Emission from the edge of the active region sees the same effective emissive area than emission from the center. Furthermore, we assume a uniform distribution of power for the different polarizations and a gaussian spectrum of the µLED from $\Lambda$ [nm] = $[425, 475]$ and a FWHM of 25nm to demonstrate that the concept is robust for a typical emission range of a µLED. $E_i(\theta, \varphi, \lambda)$ is then the electric field in the farfield region for every individual dipole $i$ at elevation $\theta$ and azimuth $\varphi$ in spherical coordinates. 
LEE$_\Gamma$ can be easily computed from $\rho_\text{LEE}$ 

\begin{align}\label{eq:LEE_gamma}
    \text{LEE}_\Gamma &= \int_0^\Gamma \d \theta \int_\Lambda \d\lambda\, \rho_\text{LEE}(\theta, \phi, \lambda)&(\text{2D})\\
    \text{LEE}_\Gamma &= \int_0^\Gamma \sin{\theta}\, \d \theta \int_0^{2\pi}\d\varphi \int_\Lambda \d\lambda\, \rho_\text{LEE}(\theta, \phi, \lambda) \hspace{.2cm} &(\text{3D})
\end{align}



\begin{figure*}[ht]
    \centering    
    \includegraphics[angle=0, trim = 0cm 0cm 0cm 0cm, clip, width = .99\textwidth]{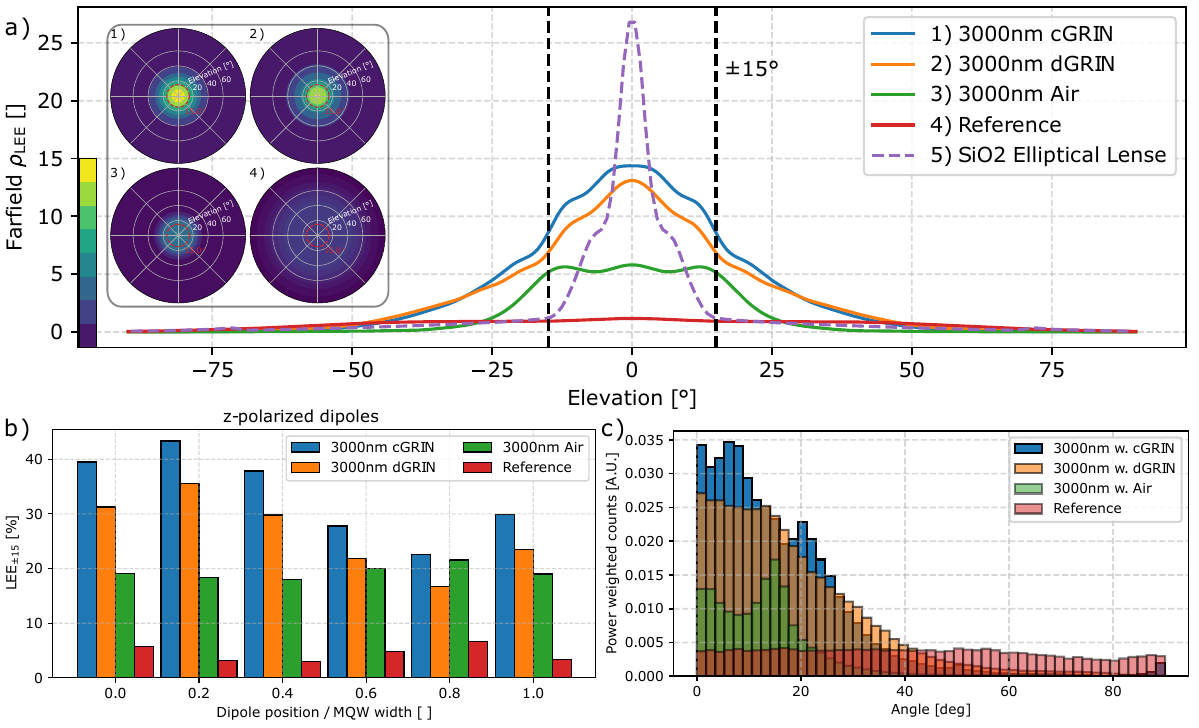}
    \caption{\textbf{| Results and investigation of the 3D cylindrical \textmu LEDs with \textmu Horn and GRIN/air filling compared with the bare reference pixel and the reference pixel with a tall SiO2 elliptical lens.} The \textmu Horn significantly improves the directionality of the µLED also in 3D. The farfield plots (a) show the LEE density in the farfield. Compared to the reference farfield (a.4) for the bare chip without outcoupling structure (see \autoref{appendix:fig:3D_reference}), using the \textmu Horn leads to a much brighter and more directional farfield. Furthermore, we added a comparison with a pixel with a large half-ellipsoidal lens with a diameter of 24\textmu m and a height of 26\textmu m, which provides increased outcoupling due to the lens material having a higher index and focusing the emission through its surface curvature. Employing such a lens can improve directionality significantly but at the cost of a much bigger structure as demonstrated in \autoref{appendix:fig:cGRIN_vs_elliptical_lens}. The enhancement via the GRIN becomes evident both in the farfield density plots a) as well as c) where the distribution of poynting vectors at the opening of the \textmu Horn is shown. The latter demonstrates that the distribution of k-vectors in the near field (right above the \textmu Horn aperture) is much more confined compared to the reference, however this can only serve as an indication of the directionality without taking the phase of the field into account. Comparing the dipole LEE$_\text{15}$ over position in the multi quantum well (MQW) for all four cases shown in b), demonstrates the advantage of the GRIN coating, in particular for the theoretical cGRIN. The LEE$_\text{15}$ remains high for the dipoles furthest from the center of the quantum well which have the highest impact on the total LEE$_\text{15}$ while substantially increasing the LEE$_\text{15}$ in the center of the MQW. \label{fig:3d_results}}
\end{figure*}


\subsubsection{2D proof of principle calculation results}

In \autoref{fig:2D_sweep_results} (a-c) and (d-f), we illustrate how LEE$_{15}$ and LEE$_{90}$ vary with the height of the \textmu Horn collimator and its sidewall angle across three distinct filling scenarios. As expected, the highest LEE$_{15}$ values are observed with the tallest \textmu Horn collimator, where more light interacts with the sidewall and gets redirected, as evidenced in \autoref{fig:2D_sweep_results} (a-c). However, the trends for LEE$_{90}$ differ. In the air-filled case, the peak LEE$_{90}$ occurs at a lower \textmu Horn height, as shown in \autoref{fig:2D_sweep_results} (d). This is because taller structures lead to increased absorption losses at the sidewall, since the filling material has the same refractive index as the ambient environment.

The trend shifts in scenarios involving dielectric fillings. In \autoref{fig:2D_sweep_results} (e) and (f), taller \textmu Horns correlate with higher LEE$_{90}$ values. This is due to the redirection of light beyond the escape cone of the \textmu Horn's filling material into the air, significantly boosting both LEE$_{90}$ and LEE$_{15}$. This behavior aligns with the Gaussian beam model depicted in \autoref{fig:concept_model_with_power_flow} (b) and (c). Notably, in \autoref{fig:2D_sweep_results} (e) and (f), the peak LEE$_{15}$ values shift to shallower \textmu Horn angles for filled \textmu Horn collimators because more light at large oblique angles must be deflected into a narrower cone within the filling material to match the 15-degree cone in air.

For the third scenario involving a discrete GRIN (Gradient Index) \textmu Horn filling, as shown in \autoref{fig:2D_sweep_results} (c) and (f), each layer step must deflect light into a specific angular range to match the 15-degree cone in air. The varying incoming light angles at the sidewalls suggest that each layer might require a unique slope. This complex case merits further investigation, but here we concentrate on elucidating the fundamental principles.

Although the results indicate that a \textmu Horn taller than 3µm would achieve better results, we intentionally omitted even taller structures for practical reasons as that would typically lead to a larger geometrical foot print which is not desirable for high resolution displays and AR/VR applications.

In this considered 2D case, compared to the reference without an outcoupling structure (LEE$_{15}$=14.9\%, see \autoref{appendix:2d_reference}), the \textmu Horn with a discrete GRIN achieves LEE$_{15}$=37.7\%, an improvement of 2.5 times. The SiO2-filled \textmu Horn achieves an LEE$_{15}$=30.4\%, while the air-filled \textmu Horn achieves an LEE$_{15}$=27.5\%.

To further understand the internal dynamics, we provide a comparison of the power flow, i.e., the time-averaged Poynting vector, incoherently averaged over the dipole positions and weighted by the µLED spectrum, across three scenarios at their highest LEE$_{15}$ values (see \autoref{fig:2D_sweep_results} (g-i). A notable similarity across the different cases is the power flow distribution along the side wall of the \textmu Horn collimator, with a visible peak at the edge in all scenarios but most pronounced in the GRIN case. This shows the continuous redirection of the light at the sidewalls due to the continuous refraction during propagation. In contrast, the power flow in the center of the outcoupling area is parallel to the normal direction. For the LEE$_{90}$, shown in \autoref{fig:2D_sweep_results} (d-f), we observe that LEE$_{90}$ is less sensitive to \textmu Horn height and angle. This is because much of the light, traditionally trapped inside the active area due to TIR at the pixel and \textmu Horn collimator interface, is refracted on the metal side walls and redirected out of the structure. The higher outcoupling generally leads to more light in the forward direction, partially explaining the higher LEE$_{15}$ for the discrete GRIN stack.

\subsubsection{3D cylindrical structures}\label{sec:3d_results}
\begin{table*}[]
    \centering
    3D validation \vspace{.1cm}
    
    \begin{tabular}{c||c|c|c}
        \hline
          Structure & LEE$_\text{15}$ $\rightarrow$ vs. Ref. & LEE$_\text{90}$ $\rightarrow$ vs. Ref. & LEE$_\text{15}$ / LEE$_\text{90}$\\
        \hline 
        Reference & \,\,\,2.3\% $\rightarrow$ 1.00x \,\,\,& 31.8\% $\rightarrow$ 1.00x\,\, & 7\%\\
        Air & 11.6\% $\rightarrow$ 5.04x \,\,\,& 27.7\% $\rightarrow$ 0.87x\,\,& 42\%\\
        SiO2 & 11.7\% $\rightarrow$ 5.09x \,\,\,& 54.1 \% $\rightarrow$ 1.70x\,\,& 22\%\\
        GaN & 9.2\% $\rightarrow$ 4.00x & 66.5\% $\rightarrow$ 2.09x& 13\% \\
        dGRIN & 19.9\% $\rightarrow$ 8.65x \,\,\,& 75.6\% $\rightarrow$ 2.38x\,\, & 26\%\\
        cGRIN & 24.6\% $\rightarrow$ 10.57x & 80.0\% $\rightarrow$ 2.52x & 31\%\\
        SiO2 lens & 11.3\% $\rightarrow$ 4.91x \,\,\,& 32.6\% $\rightarrow$ 1.03x & 35\%\\
    \end{tabular}
    \vspace{.2cm}
    \caption{Comparison of the LEE$_\Gamma$ into the $\pm15$° and $\pm90$° solid angles for the 3D validation of the 2D parameter sweep results and the improvement ratio to the reference. Its clear that the \textmu Horn significantly improves the directionality. While the \textmu Horn leads to a decrease in the LEE$_{90}$ for the air filled case, adding a filling material within the \textmu Horn has a large effect, both on the directional outcoupling as well as total outcoupling. In particular, filling the \textmu Horn with a GRIN coating improves the directionality significantly. For comparison, a large SiO2 elliptical lens was added on top of the bare reference chip which also leads to a good inhancement of directionality but at the cost of a much larger structure, shown in next to the cGRIN in \autoref{appendix:fig:cGRIN_vs_elliptical_lens}.}
    \label{tab:3D_LEE_results}
\end{table*}

Having shown the proof-of-principle calculations in 2D, we proceed to show how the principle naturally works in 3D for selected cases assuming cylindrical symmetry. Hence, the results are validated with a 3D simulation of the µLED structure for the best \textmu Horn antenna cases with $\vartheta=60^\circ$ for the GRIN and $\vartheta=72^\circ$ for the air filled case at $h=3$µm to cross check for validity in a realistic scenario. The downside of using 3D simulations is the massive increase of computational power requirements due to the scaling of FDTD with the number of mesh points $n$ which is $O(n^4)$ \cite{Schneider:2011:Understanding_the_Finite-Difference_Time-Domain_Method}.

As the considered structures have cylindrical symmetry, we only need to consider dipoles emitted along a radial cross-section, each with a proper weighting factor contribution to account for the area it represents. The weighting factor ($w_\text{i}$) is essentially described by the expression:
\begin{align}\label{eq:dipole_position_weighting}
    w_\text{i} = \frac{(r_i+\Delta r/2)^2 - (r_i-\Delta r/2)^2}{R^2}
\end{align}
where $r_i$ is the dipole position, $\Delta r$ the distance to the next dipole and $R$ the radius of the active area. A graphic explanation of the weighting is given in \autoref{appendix:dipole_position_weighting}. This position-dependent weighting indicates that, in particular, the dipoles towards the outer edges contribute relatively more to the total directional emission and light extraction efficiency, as they represent a larger area coverage of the active area. Most efforts should be directed to enhance the LEE$_\Gamma$ from dipoles located at the sides.

The results of the cylindrical 3D realizations are presented in \autoref{fig:3d_results} and compared against two reference cases: first, a bare pixel, again with 600nm diameter of GaN region, a 200nm SiO2 sidewall passivation followed by silver sidewalls (The total pixel width opening diameter comprising of GaN+SiO2 passivation is thus again 1um, see \autoref{appendix:3d_reference}). Second, considering a large SiO2-optimized half-ellipsoidal lens that is 24 times larger in diameter than the pixel's width/opening diameter, placed on top of the pixel (see \autoref{appendix:fig:cGRIN_vs_elliptical_lens}). 

The lens reference case is assessed using a hybrid approach combining wave and ray optics. Initially, we use rigorous FDTD simulations to compute the far-field emission of a \textmu LED pixel within an SiO2 ambient medium. Subsequently, we propagate this farfield as rays through a ray optics simulation tool (LightTools, see \autoref{appendix:fig:elliptical_lens_raytracing_sim}), capturing the portion of light that propagates into the ambient air within the desired cone. In the ray optics simulation, we model the farfield as emanating from a circular surface with a 1 \textmu m diameter, assuming a uniform luminance across the area. This surface is positioned at the bottom center of a half-ellipsoid SiO2 lens. The pixel and lens are further assumed to be placed on top of a silver substrate for simplicity. The 3D simulations corroborate the 2D results, demonstrating significantly higher directionality and LEE$_\text{15}$ compared to the reference case. This outcome is expected, as more light is naturally sent to oblique directions in the 3D case due to solid angle considerations, where there are simply more states/modes that can propagate at oblique polar angles.

Consequently, the impact of the \textmu Horn collimator is more pronounced in 3D than in 2D. We note, however, that the optimal geometrical parameters for the 3D structure can be significantly different from those seen in the 2D case, purely due to the fact that line dipole emission is vastly different compared to point dipole emission. Thus, by no means is the considered 3D structure here optimal. Regardless, these considered 3D realizations already exhibit a very large enhancement of the LEE$_{15}$ of up to 10 times compared to the bare pixel reference and more than twice that of the large SiO2 half-ellipsoidal lens case, as demonstrated in \autoref{fig:3d_results} a). The numerical results are shown in \autoref{tab:3D_LEE_results}, which demonstrates that adding an appropriate \textmu Horn increases the directionality for all cases significantly.

In addition to the 3D realization of the dGRIN, we examined the performance of a continuous GRIN (cGRIN) with the same geometric parameters as the dGRIN shown in \autoref{fig:3d_results}a) (1) and (2). As anticipated, both directionality and LEE$_{90}$ are further enhanced due to the seamless refractive index transition between the semiconductor core of the µLED and the ambient medium. A comparison of the absolute values is provided in \autoref{tab:3D_LEE_results}. The reference µLED exhibits low directionality but a comparable LEE$_{90}$ to the \textmu Horn filled with air.

We caution the reader that far-field cross sections can be misleading due to the $\sin{\theta}$ factor in \autoref{eq:LEE_gamma}. This is evident when examining the spherical far-field plots shown in \autoref{fig:3d_results}a) or the $\sin{\theta}$ corrected far-field cross-sections shown in \autoref{appendix:fig:sin_theta_corrected_farfields}.

Furthermore, a more detailed investigation into the individual dipole efficiencies reveals the reason for the increased absolute efficiency of the GRIN. By resolving the far-field angle-dependent efficiency of the individual dipoles, shown in \autoref{fig:3d_results}b), we can observe multiple contributions. For all cases with the \textmu Horn, the LEE$_{15}$ of all individual dipoles is increased, hence the much more directional emission and overall higher LEE$_{15}$. Interestingly, the dipole LEE$_{15}$ of the reference and air-filled \textmu Horn remain relatively stable over the quantum well, while the d- and cGRIN experience a decrease in LEE$_{15}$. This is particularly important because, due to the weighting factor of the dipole position, the dipoles towards the outer edge contribute more to the absolute LEE$_\Gamma$. However, the individual LEE$_{15}$ remains higher for all dipoles by adding a GRIN filling, which results in an absolute improvement of the LEE$_{15}$ for the µLED. Here, we show the efficiency for z-dipoles only. The LEE$_{15}$ efficiencies for x- and y-polarization are shown in \autoref{appendix:fig:LEE15_polarization_efficiencies} and for LEE$_{90}$ in \autoref{appendix:fig:LEE90_polarization_efficiencies}.

While the analysis of the individual dipole LEE indicates that the light extraction efficiency throughout the quantum well was improved, we can also see the impact of the \textmu Horn collimator in \autoref{fig:3d_results}c). As discussed in \autoref{sec:Effect of the Horn antenna on µLED emission}, the presence of the \textmu Horn reduces the width of the angular distribution significantly compared to the almost uniform distribution from the bare reference pixel. Compared to all cases, the cGRIN structure focuses the largest proportion of light into the $\pm15^\circ$ solid angle. It is also visible that the air-filled \textmu Horn reduces the width of the distribution even further but does not achieve the same amount of total power as the GRIN versions. This is a consequence of the lower overall extraction efficiency and demonstrates the trade-off between increasing directionality at the expense of light extraction efficiency.

\begin{figure*}[h]
  \centering
  \begin{subfigure}[h]{0.49\textwidth}
    \includegraphics[angle=0, trim=0cm 0cm 0cm 0cm, clip, height=6cm]{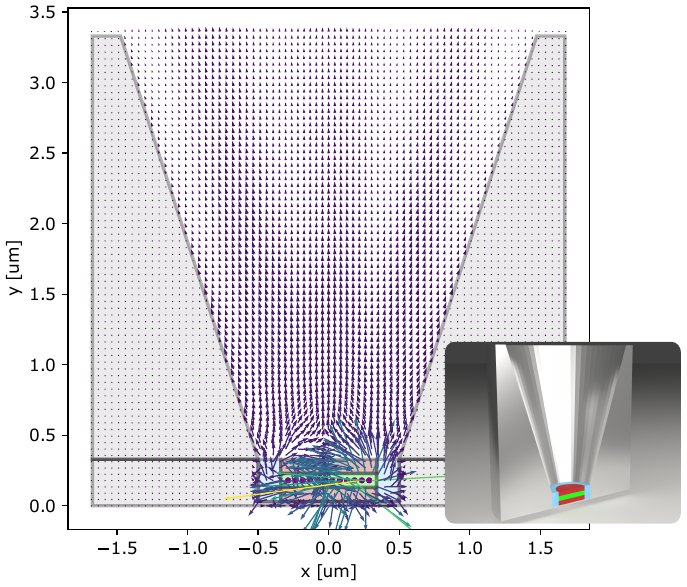}
    \caption{\textbf{| Asymmetric power flow through the \textmu Horn filled with air.} We show the incoherent, weighted superposition of only half of the active region. Due to the high refractive index contrast, most of the power is trapped within the µLED core resulting in the low LEE$_{90}$. The steeper angle of the \textmu Horn compared to the cGRIN shown in \autoref{fig:cGRIN_asymmetric_powerflow} leads to a more narrow near field distribution which leads to a higher fraction of LEE$_{15}$ / LEE$_{90}$. However, the absolute LEE$_{15}$ is still well below the LEE$_{15}$ for the cGRIN, as shown in \autoref{tab:3D_LEE_results}.}
    \label{fig:air_asymmetric_powerflow}
  \end{subfigure}
  \hfill
  \begin{subfigure}[h]{0.49\textwidth}
    \includegraphics[angle=0, trim=0cm 0cm 0cm 0cm, clip, height=6cm]{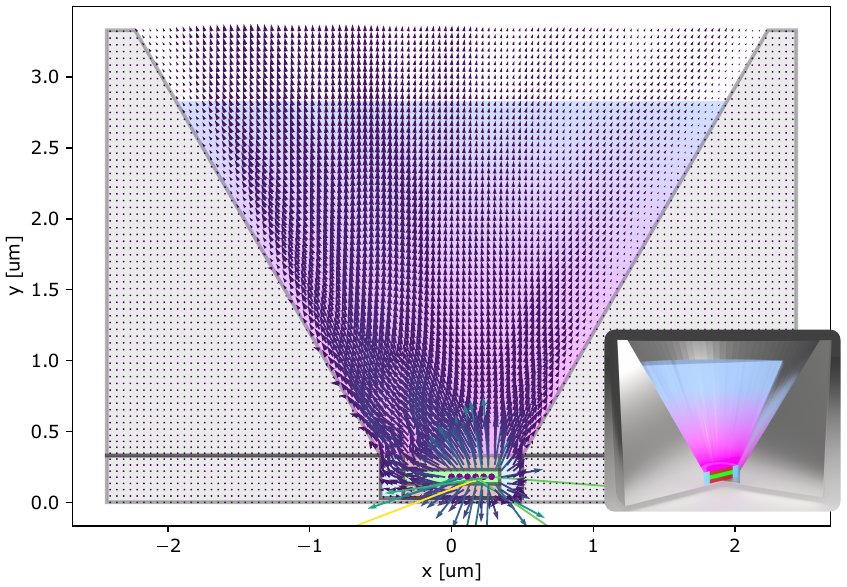}
    \caption{\textbf{| Asymmetric power flow through the cGRIN \textmu Horn.} We show the incoherent, weighted superposition of only half of the active region. Due to the index match between the cGRIN and the active region, most of the power can escape the µLED core. The power is subsequently reflected at the sidewall on the opposite site of the emission region. After the interception of the emission at the sidewall, a large part of the radiation is redirected and exits the µLED opening close to perpendicular.}
    \label{fig:cGRIN_asymmetric_powerflow}
  \end{subfigure}
  \caption{Comparison of asymmetric power flow in \textmu Horns filled with air and cGRIN.}
\end{figure*}

The effect of lower extraction efficiency is further exemplified when comparing the asymmetric power flows of the \textmu Horn filled with air in \autoref{fig:air_asymmetric_powerflow}) with those of the \textmu Horn with the cGRIN shown in \autoref{fig:cGRIN_asymmetric_powerflow}). The significant refractive index difference between the active area and the air inside the \textmu Horn results in much lower outcoupling efficiency, which effectively traps most of the light at oblique angles inside the active region. Instead of interacting with the \textmu Horn, this light experiences significant absorption, leading to the lower overall extraction efficiency that occurs in the absence of the GRIN filling within the horn collimator.



\section{Conclusion}
\label{sec:5_conclusion}



We have shown the impact of a Horn collimator with different material fillings on the emission characteristics of a simplified \textmu LED model. The proposed GRIN-filled Horn significantly enhances both the total light extraction efficiency within the $\pm$15° cone by an order of magnitude compared to the bare \textmu LED pixel and more than doubles it compared to the case of a \textmu LED pixel with a vastly large half-ellipsoidal SiO2 lens, whose diameter is 40 times the active area, as confirmed by full 3D FDTD simulations. Depending on the choice of material that fills the Horn, the optimal opening angle is shifted, which is a consequence of the trade-off between maximizing the directionality and the total extraction efficiency. Regardless, the enhancement is achieved with a comparatively small geometrical footprint, making it ideal for many display and augmented reality (AR) applications \cite{Xiong:2021:Augmented_reality_and_virtual_reality_displays:_emerging_technologies_and_future_perspectives, Kang:2024:Advances_in_display_technology:_augmented_reality_virtual_reality_quantum_dot-based_light-emitting_diodes_and_organic_light-emitting_diodes} or optical communications \cite{Lu:2022:High-speed_visible_light_communication_based_on_micro-LED:_A_technology_with_wide_applications_in_next_generation_communication, Singh:2020:Micro-LED_as_a_Promising_Candidate_for_High-Speed_Visible_Light_Communication}.

This improvement can be attributed to a combination of factors: the enhancement of outcoupling efficiency over the active region, driven by the refractive index / impedance matching of the GRIN with the active region, and the improved redirection of emission at the Horn collimator. This results in a narrow angular distribution in the near field.

The investigation presented in this article identifies the key contributions to achieving ultra-high LEE and directional \textmu LEDs, though it is not exhaustive. Significant improvements and geometrical optimizations are still possible, which could further enhance directionality and outcoupling efficiency \cite{Vuckovic:2019:Nanophotonic_Inverse_Design_with_SPINS:_Software_Architecture_and_Practical_Considerations, Wang:2019:Velocity_field_level-set_method_for_topological_shape_optimization_using_freely_distributed_design_variables, Molesky:2018:Inverse_design_in_nanophotonics, Luce:2024:Merging_automatic_differentiation_and_the_adjoint_method_for_photonic_inverse_design, Augenstein:2020:Inverse_Design_of_Nanophotonic_Devices_with_Structural_Integrity}. Specifically, geometrical designs that maintain high coupling efficiency while producing a narrower angular distribution could result in even higher LEE$_\Gamma$ than demonstrated in this work. 

Possible geometrical optimizations include the following areas of investigation:
\begin{itemize}
    \item We assumes a particularly simple \textmu LED core geometry but the core and the interaction of the dipoles with the surrounding has a large effect on the outcoupling efficiency and directionality \cite{Vogl:2024:Optical_characteristics_of_thin_film-based_InGaN_micro-LED_arrays:_a_study_on_size_effect_and_far_field_behavior}. Altering the core therefore could enhance the LEE$_\Gamma$.
    \item We assumed a linear decrease in refractive index for the GRIN, which is straightforward. However, it remains unclear if a different refractive index distribution might be beneficial such as a quadratically decreasing profile which is often used in optical gradient index fibers. 
    \item The Horn angle was assumed to be constant throughout this article although the results of this work already show that the optimal Horn angles depends on the filling material/refractive index. Investigating a variable angle Horn profile might yield significant improvements over the current linear profile. 
\end{itemize}

\subsection*{Authors contribution}
The concept of this paper was jointly envisioned by A. Abass, R. Alaee, and A. Luce. A. Luce carried out the implementation, study, and analysis. The interpretation of the results was a collaborative effort among A. Abass, R. Alaee, and A. Luce. A. Luce wrote the manuscript, with guidance, corrections and suggestions from A. Abass and R. Alaee.

\subsection*{Acknowledgments}
We thank the supporters of this work, particularly Chuan Hong Lau (former ams OSRAM), Fabian Knorr (ams OSRAM), and Harald Laux (ams OSRAM) for his organizational support.

\subsection*{Publication Funding}
Funding from the German Federal Ministry of Economic Affairs and Climate Action (BMWK) and the Bavarian State Ministry of Economic Affairs and Media, Energy and Technology within the IPCEI-ME/CT “OptoSure (GA: 16IPCEI221) is gratefully acknowledged.

\bibliographystyle{unsrt}  
\bibliography{0_references}  

\setcounter{section}{0}
\setcounter{figure}{0} 
\renewcommand\thefigure{\thesection.\arabic{figure}}   
\renewcommand\thesection{\Alph{section}}
\renewcommand\thesubsection{\thesection.\arabic{subsection}} 
\section{Appendix}
\label{sec:appendix}

\subsection{Huygens source diffraction}\label{appendix:huygens_source_diffraction}
As discussed in \autoref{sec:theory}, one can think of the opening area of a \textmu LED as an aperture. The width of the incident angular spectrum of the light and the width of the aperture of course play a significant role for the farfield of the aperture and therefore for the \textmu LED. In \autoref{appendix:fig:huygens_source_diffraction}, the illumination of an aperture with width of 1\textmu m is shown. Unsurprisingly, a broader angular distribution leads to a much broader intensity distribution. Here, we assume a uniform distribution of incident angles given by the width $\Delta \theta$ (y-axis). For a specific distribution $\Delta \theta_i$, the incoherent intensity in the farfield is simply the integral over the incident angles $I(\vartheta) = \int_{-\Delta \theta}^{\Delta \theta}\d \theta I(\vartheta, \theta)$. µLEDs are generally limited by the diffraction of the emission in the outcoupling region, which can be considered to be an aperture with the appropriate cross section shape. Therefore, reducing the angular distribution and increasing the aperture diameter/width leads to stronger directionality of the farfield. An explicit derivation of the farfield intensity for a square aperture can be found in \cite{Biswas:2021:Explicit_derivation_of_the_Fraunhofer_diffraction_formula_for_oblique_incidence}.

\begin{figure}[ht]
  \centering
  \includegraphics[angle=0, trim = 0cm 0cm 0cm 0cm, clip, width = .99\columnwidth]{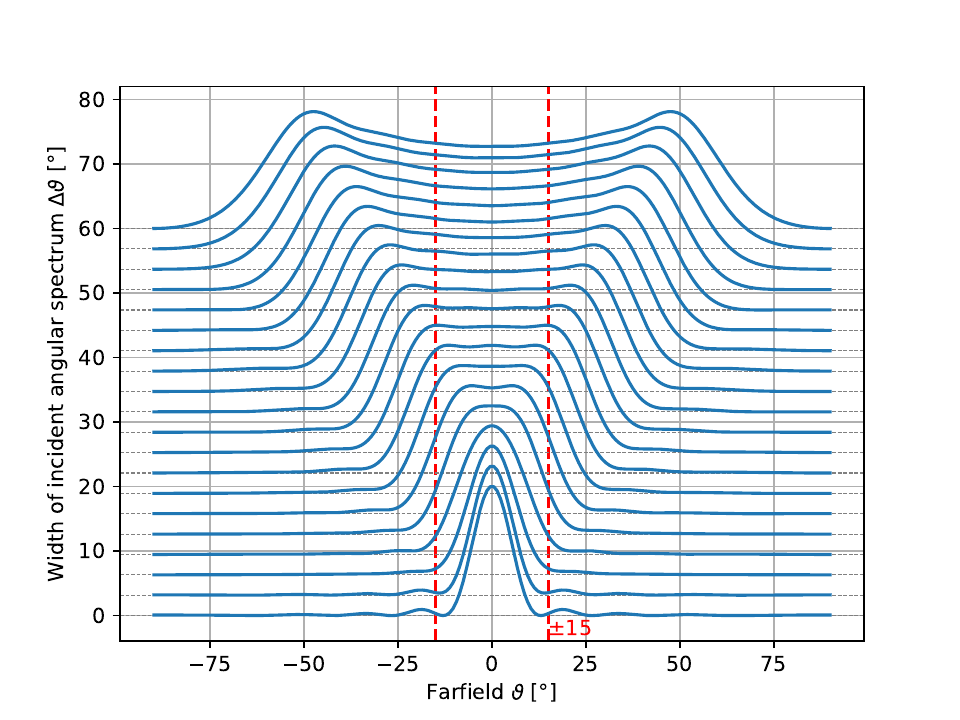}
  \caption{\textbf{| Farfield intensity distribution for an incoherent superposition of plane waves with an varying near field angular distribution width.} The plane waves illuminate a line aperture with a width $d= 1$\textmu m. A larger width of the incident angular distribution leads to a much broader farfield which impacts the directionality of \textmu LEDs. \label{appendix:fig:huygens_source_diffraction}}
\end{figure}

\subsection{2D Reference}\label{appendix:2d_reference}
\begin{figure}[ht]
  \centering
  \includegraphics[angle=0, trim = 0cm 0cm 0cm 0cm, clip, width = .99\columnwidth]{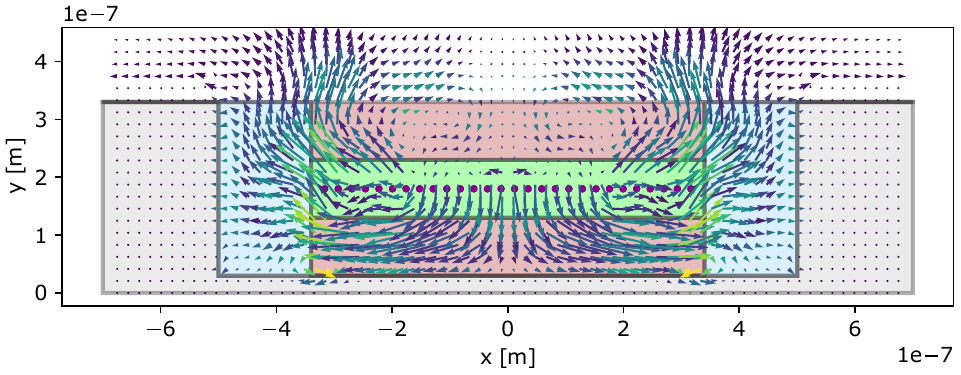}
  \caption{\textbf{| \textmu LED crosssection power flow of the bare pixel reference model, simulated in 2D.} \label{appendix:fig:2d_reference}}
\end{figure}

In order to compare the results from the analysis, performed in \autoref{sec:Effect of the Horn antenna on µLED emission}, we computed a bare pixel reference in 2D. In general, performing an FDTD simulation in 2D means to assume translational symmetry outside of the simulation plane which results in the dipoles behaving as a coherent line emitter in the direction of the translational symmetry. This introduces an error in the simulation results and hence, the 2D results must be interpreted with caution. The power flow of inside the reference is shown in \autoref{appendix:fig:2d_reference}. The total LEE$_{15}$ = 14.9\% and LEE$_{90}$ = 55.9\%. Compared to the results in 3D, shown in \autoref{appendix:3d_reference}, the 2D simulation results show a significantly better performance. This is likely an effect of the different density of optical states in 2D compared to 3D.



\subsection{Dipole position weighting}
\label{appendix:dipole_position_weighting}
\begin{figure}[ht]
    \centering
    \begin{subfigure}[b]{0.44\columnwidth}
        \centering
        \includegraphics[width=\textwidth]{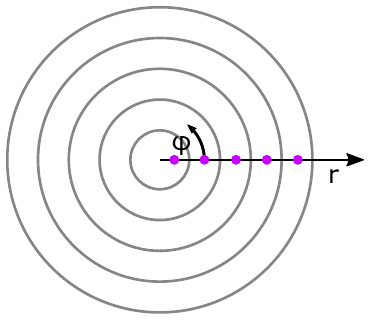}
    \end{subfigure}
    \hfill
    \begin{subfigure}[b]{0.55\columnwidth}
        \centering
        \includegraphics[width=\textwidth]{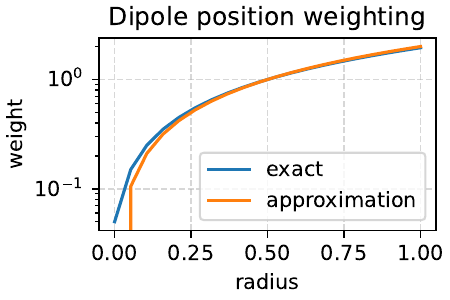}
    \end{subfigure}
    \caption{\textbf{| Visual depiction of the dipole position weighting and comparison between the exact and approximated weighting factors.} The weighting factor is necessary because an emitting dipole at larger radius r from the center of the active region has a larger effective emissive area as shown in (a). The exact weighting used is described in \autoref{eq:dipole_position_weighting} while the approximated weighting factor is $w_i' = 2 \cdot r_i / R^2$. The latter is often used in literature but deviates from the exact surface area weighting factor for small r and a small number of simulated dipoles. \label{appendix:fig:dipole_position_weighing}}
\end{figure}
Large 3D FDTD simulations can be very time consuming, especially on high resolution which is necessary to obtain accurate results. Thus, its very beneficial to save as many simulation runs as possible. For µLEDs, one needs to perform multiple simulations in parallel to receive the correct system behavior because of the coherence properties of the emitters in LEDs. In short, every point within the quantum wells of the active region acts as an independent emitter which emits incoherently from all other emitters. Hence, in order to get an accurate result without spurious correlations between sources, its necessary to perform a simulation for every dipole individually and superposition the optical power only during post processing. Solving so many dipoles quickly becomes infeasible. In order to reduce the computational effort, an obvious way is to employ the symmetry of the problem. We consider a cylindrical µLED and therefore, the fields produced from any dipole is also rotationally symmetric around the center axis of the µLED. We can then reduce the necessary amount of dipoles by only solving dipoles on a radius from the center and assuming that the fields emitted by different dipoles within $\pm \Delta r$ around the dipole are similar. From \autoref{appendix:fig:dipole_position_weighing}, it is clear that dipoles closer to the center have a much smaller effective emissive area and therefore contribute less to the total farfield with the contribution proportional to the surface area of the circular rings which hold the dipole. The equation giving the exact weighting factors is shown in \autoref{eq:dipole_position_weighting}. For completeness, in other works a different weighting factor is sometimes employed which approximates the surface area weighing by $w_i' = 2 \cdot r_i / R^2$. Both weightings are almost identical for small $\Delta r$. However, since the number of dipole positions is rather small, we chose to use the exact weighting proportional to the effective surface area of the dipoles, see \autoref{appendix:fig:dipole_position_weighing}

\subsection{3D References}\label{appendix:3d_reference}
The results from the reference geometry without Horn are shown in \autoref{appendix:fig:3D_reference}. Compared to the reference results in 2D, we observe very different absolute values of LEE$_{15}$ and LEE$_{90}$. Additionally, we show the impact of the additional weighting factors on the power flow and the far fields. First of all, the wavelength weighting with a gaussian leads to a rescaling of the farfields of different wavelenghts. The impact of the dipole position weighing can be seen in the power flow. The unweighted case shows more pronounced emission from the center, a consequence of the generally higher outcoupling efficiency of the central dipoles. By introducing the position dependent weighting factor, detailed in \autoref{appendix:dipole_position_weighting}, we see that the emission of the outer dipoles becomes more important. Due to the generally lower outcoupling efficiency of the outer dipoles, see \autoref{fig:3d_results},  \autoref{appendix:fig:LEE15_polarization_efficiencies} and \autoref{appendix:fig:LEE90_polarization_efficiencies}, the total LEE$_{15}$ and LEE$_{90}$ is reduced in the weighted case. 

\begin{figure*}[ht]
  \centering
  \includegraphics[angle=0, trim = 0cm 0cm 0cm 0cm, clip, width = .99\textwidth]{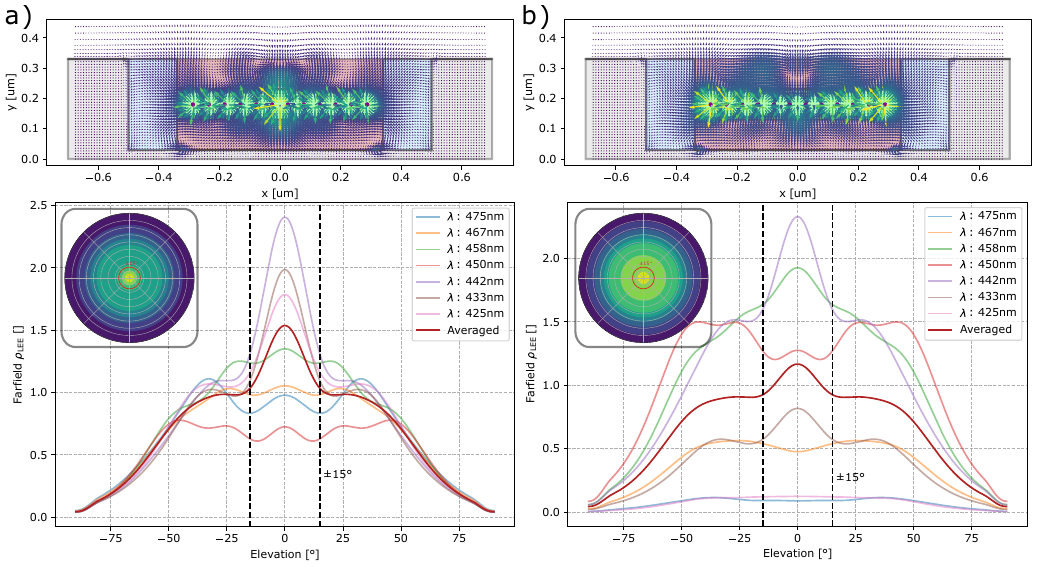}
  \caption{
      \textbf{| Power flow and farfields of the reference a) unweighted and b) weighted.} The impact of the weighting factor is immediately visible from the changing farfield density of the spectrum and the power distribution around the quantum wells. In particular, the radiated power by the dipoles at the outer edges is enhanced while the center dipole radiates less intensively. Combined with the lower LEE of the outer dipoles, it is clear why the position dependent weighting has a detrimental effect of the LEE$_{15}$ and LEE$_{90}$. For the unweighted case a), we find that LEE$_{15} = 2.7\%$ and LEE$_{90} = 32.9\%$  and weighted for case b) LEE$_{15} = 2.3\%$ and LEE$_{90} = 31.8\%$. \label{appendix:fig:3D_reference}
  }
  \vspace{-.5cm}
\end{figure*}

In order to make a comparison with a \textmu lens that could be used alternatively to the Horn and GRIN structure, a small study was performed to find the best height of an half-ellipsoidal SiO2 lens considering the LEE$_{15}$. For this, an ellipsoidal lens with fixed diameter d=26\textmu m was placed on top of the reference chip shown in \autoref{appendix:fig:elliptical_lens_raytracing_sim} and the improvement of directionality was analyzed. The results are shown in \autoref{appendix:tab:Sio2_sweep} where the best LEE$_{15}$ was achieved with a height of h=24\textmu m. In contrast to the cGRIN structure, the \textmu lens is significantly larger as shown in \autoref{appendix:fig:cGRIN_vs_elliptical_lens}.

\begin{figure}[ht]
  \centering
  \includegraphics[angle=0, trim = 0cm 0cm 0cm 0cm, clip, width = .89\columnwidth]{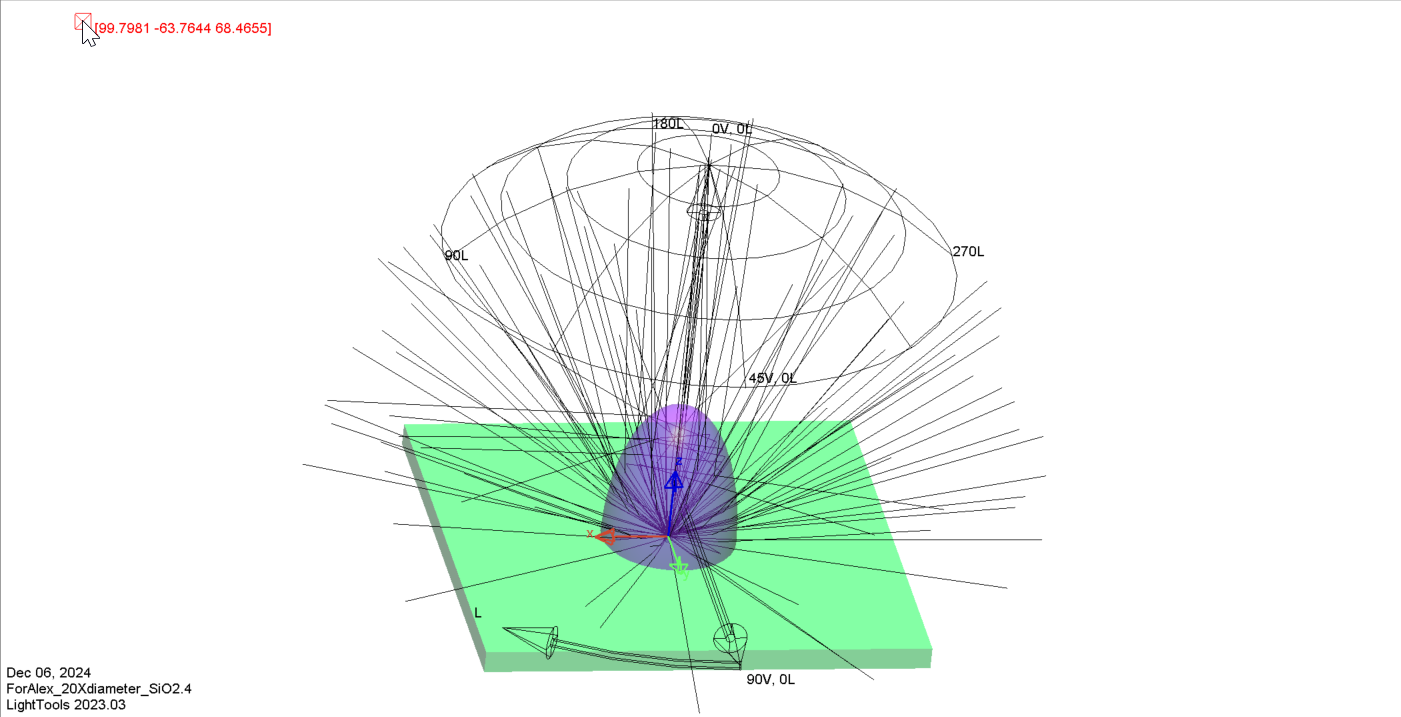}
  \caption{\textbf{| Depiction of the \textmu lens for raytracing.} The raytracing was performed via a hybrid simulation where the near-field of the bare reference pixel was recored in a SiO2 ambient medium and then transferred to a raytracing software to simulate the effects of the \textmu lens. The LEE$_{15}$ and LEE$_{45}$ are shown in \autoref{appendix:tab:Sio2_sweep}. \label{appendix:fig:elliptical_lens_raytracing_sim}}
\end{figure}

\begin{table}
\centering
SiO2 lens height vs LEE$_{15}$
    \begin{tabular}{c||c|c}
        \hline
        \textmu lens height {\textmu m]}  & LEE$_{15}$ [\%] & LEE$_{45}$ [\%]\\
        \hline
        12 & 2.1291 & 16.249 \\
        14 & 2.9205 & 19.1114 \\
        16 & 4.1521 & 20.7646 \\
        18 & 5.8591 & 21.8332 \\
        20 & 7.8325 & 22.5493 \\
        22 & 9.6361 & 22.936 \\
        24 & 10.8941 & 22.9855 \\
        26 & 11.3477 & 22.5142 \\
        28 & 7.2165 & 18.9926 \\
        30 & 4.8843 & 15.4648 \\
        32 & 3.6864 & 13.6228 \\
    \hline
    \end{tabular}
    \vspace{0.3cm}
    \caption{ 
        \textbf{| Tested heights of SiO2 half-ellipse lens and their respective LEE$_{15}$.} With a fixed base diameter of 24\textmu m and the emission calculated from the bare reference pixel in \autoref{appendix:fig:3D_reference}, the lens with h = 26\textmu m shows the best LEE$_{15}$ and is used for comparison in \autoref{fig:3d_results}.
    }
    \label{appendix:tab:Sio2_sweep}
\end{table}

\begin{figure}[ht]
  \centering
  \includegraphics[angle=0, trim = 0cm 0cm 0cm 0cm, clip, width = .89\columnwidth]{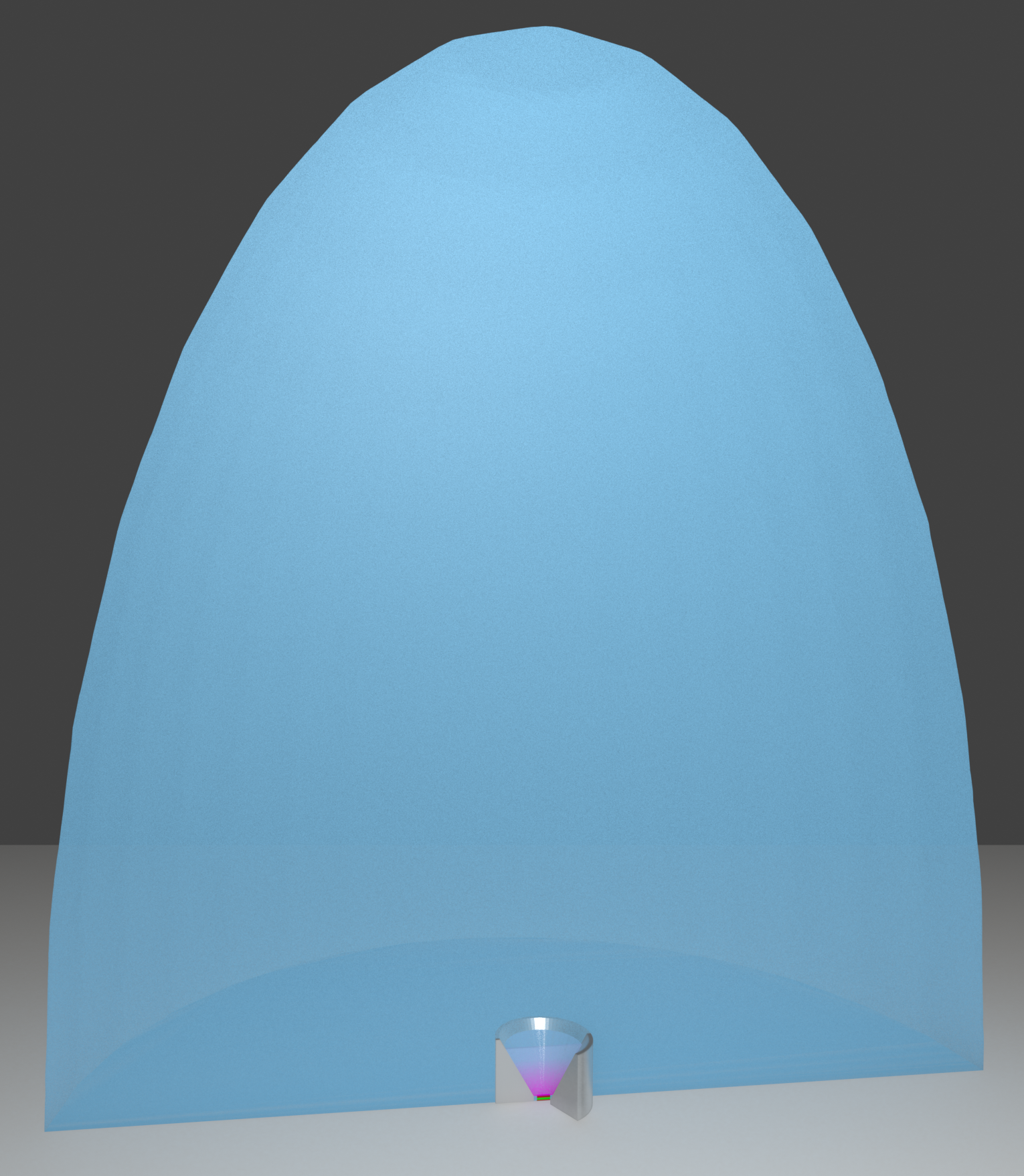}
  \caption{\textbf{| Comparison of the cGRIN emitter and an half-ellipse lens with high directionality.} The corresponding farfields are shown in \autoref{fig:3d_results} a.1) for the cGRIN and a.5) for the elliptical lens. Although the elliptical lens achieves good directionality, the LEE15 is still significantly smaller for the elliptical SiO2 lens compared to the cGRIN. Furthermore, the required size of the device is by far to large for AR/VR applications. \label{appendix:fig:cGRIN_vs_elliptical_lens}}
\end{figure}

\subsection{sin($\theta$) corrected farfield comparison}
Interpreting the farfield results, particularly as a cross-section over the elevation of the farfield, can be misleading because it is easy to omit the $\sin \theta$ factor of the differential surface element in spherical coordinates. To make the power distribution in the farfield more visually clear, it is useful to show the farfield with the $\sin \theta$ correction factor, as illustrated in \autoref{appendix:fig:sin_theta_corrected_farfields}. This correction provides a better indication of how the power is actually distributed in the farfield compared to naive farfield cross-sections that do not account for the differential surface area in spherical coordinates. Although a much larger fraction of the light is within the $\pm 15^\circ$ cone for the d- and cGRIN, most of the power is still radiated into a larger solid angle, indicating further potential for improvement.

\begin{figure}[!ht]
  \centering
  \includegraphics[angle=0, trim = 0cm 0cm 0cm 0cm, clip, width = .99\columnwidth]{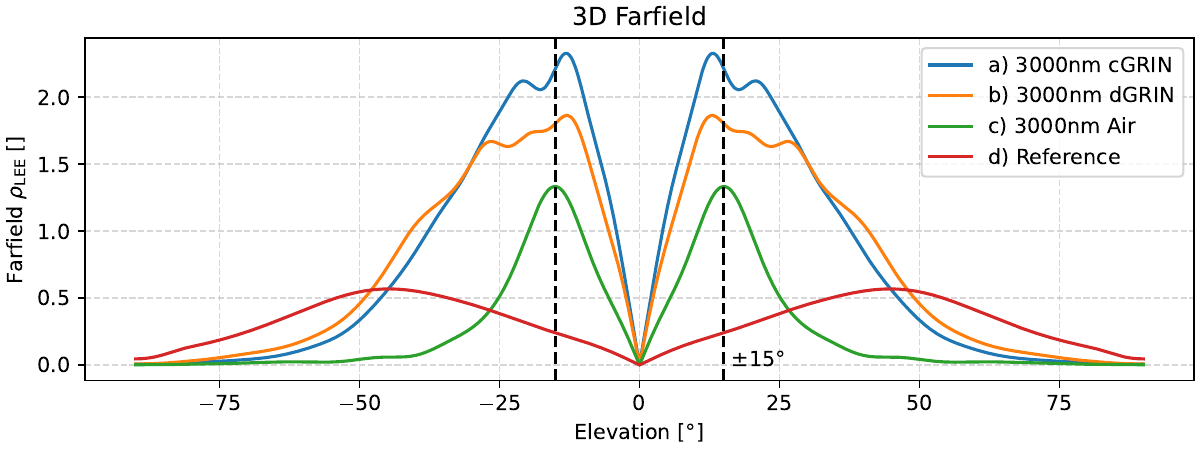}
  \caption{\textbf{| sin $\mathbf{\theta}$ corrected farfields of the 3D cylindrical models.} The uncorredcted farfields are shown in \autoref{fig:3d_results}a). The correction helps to immediately visualize to which solid angle the power is going which might be misleading when only considering the crosssection of the farfield. \label{appendix:fig:sin_theta_corrected_farfields}}
\end{figure}

\subsection{Polarization efficiencies}\label{appendix:polarization_efficiencies}

For completeness, we show the additional dipole position-dependent efficiency plots of LEE$_{15}$ for x-, and y-oriented dipoles in \autoref{appendix:fig:LEE15_polarization_efficiencies} and LEE$_{90}$ for all polarizations in \autoref{appendix:fig:LEE90_polarization_efficiencies}. We denote the dipole axis as reference for the orientation. Since we orient the \textmu LEDs such that the light is emitted in y-direction, the y-polarized dipoles primarily emit \textit{in-plane} while the x- and z-polarized dipoles emit \textit{out-of-plane}. This explains why the efficiencies for the y-polarized dipoles remains typically lower than the x- and z-polarized dipoles. 

\begin{figure}[ht]
  \centering
  \includegraphics[angle=0, trim = 0cm 0cm 0cm 0cm, clip, width = .99\columnwidth]{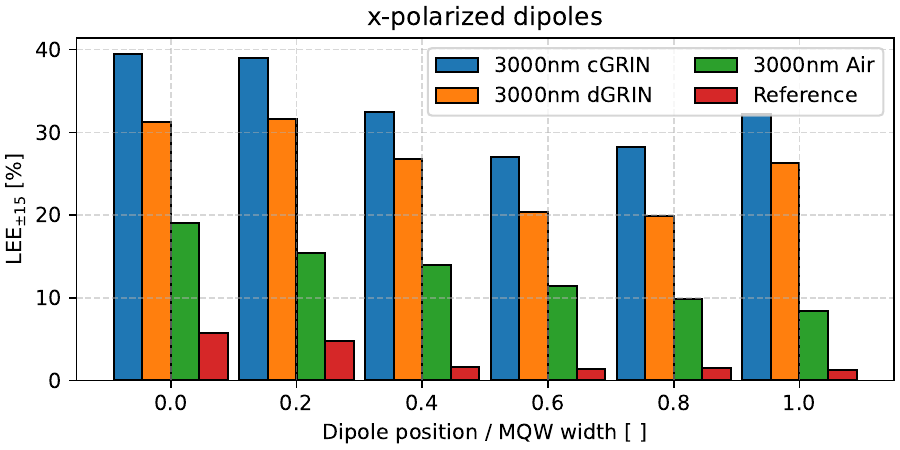}
  \includegraphics[angle=0, trim = 0cm 0cm 0cm 0cm, clip, width = .99\columnwidth]{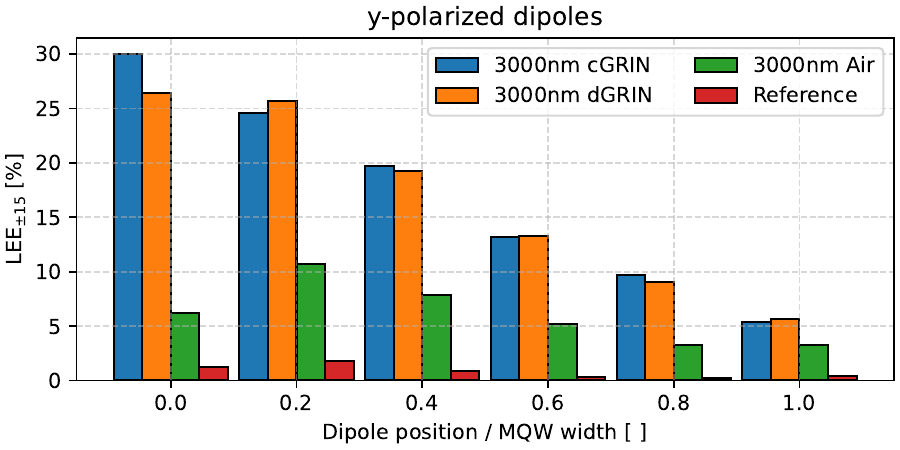}
  \caption{ 
      \textbf{| Dipole position dependent LEE$_{15}$ for x- and y-polarized dipoles.} Just like the efficiencies shown in \autoref{fig:3d_results}b), the dipole efficiencies for the GRIN structures is significantly higher for air in in particular compared to the bare reference pixel
      \label{appendix:fig:LEE15_polarization_efficiencies}
  }
\end{figure}

\begin{figure}[ht]
  \centering
  \includegraphics[angle=0, trim = 0cm 0cm 0cm 0cm, clip, width = .99\columnwidth]{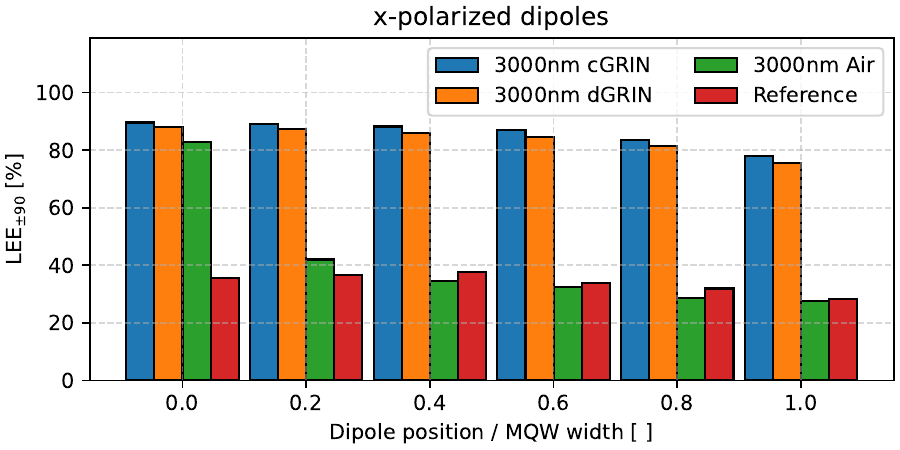}
  \includegraphics[angle=0, trim = 0cm 0cm 0cm 0cm, clip, width = .99\columnwidth]{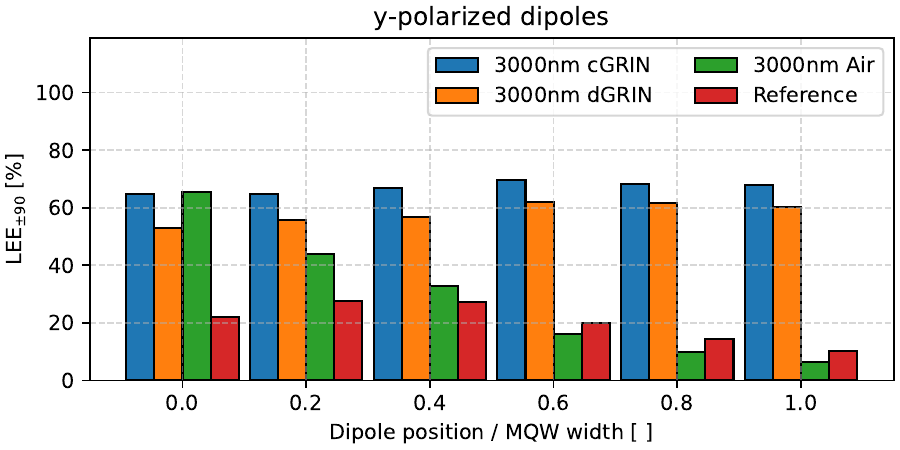}
  \includegraphics[angle=0, trim = 0cm 0cm 0cm 0cm, clip, width = .99\columnwidth]{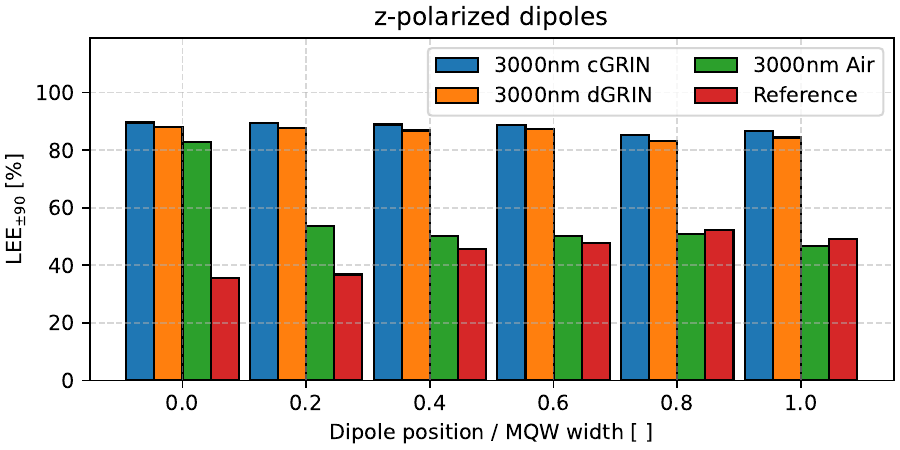}
  \caption{
      \textbf{| Dipole position dependent LEE$_{90}$ for x-, y-, and z-polarized dipoles.} Just like the LEE$_{15}$ dipole efficiencies, the GRIN structure greatly enhances total light extraction efficiency as well. \label{appendix:fig:LEE90_polarization_efficiencies}
  }
\end{figure}

\end{document}